\documentclass[aps,groupedaddress,amsmath,reprint,longbibliography]{revtex4-1}

\bibpunct{}{}{,}{s}{}{}

\usepackage{dcolumn} 
\usepackage{graphicx} 
\usepackage{color}

 \newcommand{\figwide}{7.8cm}

\begin{document}

\title{iPad Screencasts as Evidence of Accountable Disciplinary Knowledge} 
\date{\today}

\author{Eleanor C. Sayre} 
\email{esayre@ksu.edu}
\affiliation{Department of Physics, Kansas State University, Manhattan, Kansas 66506}

\author{Claudia Fracchiolla} 
\affiliation{Department of Physics, University of Colorado, Boulder 80302}

\author{Ben Van Dusen}
\affiliation{Science Education, California State University Chico, Chico, CA 95929}

\begin{abstract} 
We define knowledge as facts, information, and skills acquired by a person through experience or education.  In turn, Disciplinary Knowledge is the knowledge related to a specific field or discipline, such as physics.  Accountable Disciplinary Knowledge (ADK) is ``what counts" as important to a community or discipline: the kinds of activities that participants engage in, the problems they value and solve, and methods for solving those problems\cite{Stevens2008}. Because ADK is determined on a community basis, the details of "what counts" will vary with the norms and expectations of the community.
In this paper, we use the ADK framework to investigate "what counts" as solving problems well in a high school physics classroom from the perspective of the students.   
\end{abstract}


\maketitle

%%%%%%%%%%%%%%%%%%%%%%%%%%%%%%%%%%%%%%%%%%%% 
%% MAINMATTER 
%%%%%%%%%%%%%%%%%%%%%%%%%%%%%%%%%%%%%%%%%%%%

\section{Introduction}

%%%%%%%%%%%%%%%%%%%%%%%%%%%%%%

Problem solving is an integral part of most physics classes\cite{Maloney1993a}. There is ample research on the benefits of developing problem-solving skills (e.g. \cite{Hsu2004}), and even a whole line of research on problem-based learning . Unfortunately, for the most part physics classes in high school and introductory physics in college still today heavily rely on end-of-chapter exercises. Research has shown that proficiency at solving end-of-chapter exercises does not promote proficiency at solving conceptual problems, although proficiency at conceptual problem solving promotes proficiency at exercise solving\cite{Heller1992a,Reif1982}.  Scaffolding students to use prescribed problem solving strategies is helpful\cite{Heller1992a, Etkina2010} for both kinds of problems.

Students prefer to frame problems narrowly\cite{Irving2013framing,Engle2006}, eschewing real-world connections and connections to deeper principles\cite{Schoenfeld1999} in favor of merely getting through the problem at hand.  Why shouldn't they? They have a large number of problems to solve every week, and if their instructor uses an online grading management system like WebAssign or MasteringPhysics, they are graded only on answers, not framing or principles.  It's time-consuming and grade-producing. This relates to the concept of utility value\cite{Hulleman2008}: the degree to which a task is perceived as being relevant beyond the immediate situation.  In this paper, we take up the question ``what is the utility of exercises?'' from the lens of what solving exercises teaches students about learning physics.

%%%%%%%%%%%%%%%%
Recent research has shown that how learning is framed can influence what strategies students use, how much they enjoy learning activities, and ultimately what they transfer\cite{Hammer2000, Engle2006}. 
%Research on epistemological framing studies show how the perception of students affects what tools and skills they believe are needed \cite{Irwing16}. 
In addition to the technical content that students learn while they solve problems, students also learn cultural content about what physicists value.  For example, physicists value reducing complicated scenarios into simpler ones (e.g. the spherical cow; the frictionless table; the massless string; the point particle).\cite{Sayre2015BESM} Physicists value problem solving, particularly in mathematical ways.  However, the ways in which physicists value problem solving -- and the kinds of problems we value as physicists -- depend on setting of the problem and the people doing the solving. 

%%%%%%%%%%%%%%%%%%%%%%%%%
To articulate the differences between physicists-solving-problems and physics-students-solving-problems, we turn to accountable disciplinary knowledge (ADK). 
From this perspective, knowledge is facts, information, and skills acquired by a person through experience or education. In turn, Disciplinary Knowledge is the knowledge related to a specific field or discipline, such as physics. Accountable Disciplinary Knowledge (ADK)\cite{Stevens2008, Irving2014AdLab} is ``what counts'' as important to a community or discipline: the kinds of activities that participants engage in, the problems they value and solve, and methods for solving those problems. Demonstrating one's ADK is central to being accepted as a member of a community of practice\cite{Lave1991, Wenger1998}. Because a community's ADK is continually being rendered and refined by its members, the norms and expectations of the community (or ``what counts'') will vary across both communities and across time\cite{Irving2014AdLab}. 

In communities of professional physicists, ADK includes posing, solving, and evaluating novel research questions.  Often the same person (or research group) who poses the question will be the one to answer it. In contrast, introductory and high school physics classroom communities' ADK often include solving short problems with answers in the back of the textbook. In these classroom settings teachers are typically responsible for assigning problems, and students are responsible for performing them. Problem solving -- in either community -- is an integral part of their ADK, but the details of what counts as a problem, how it should be solved, and what is important are different. As the physics classroom is seen as the pathway to preparing and inducing new members to the professional physics community, examining the disconnect between these versions of problem solving may provide insight into the nature of the broader populations' lack of scientific literacy\cite{PCAST2012,GatheringStorm}.

In this paper, we use the ADK framework to investigate ``what counts" as solving problems well in a high school physics classroom from the perspective of the students.  We present the case of ``Stu'', a high school student presenting an end-of-chapter-style problem to his peers, to argue that students' ADK around problem solving centers around use of correct procedure to produce correct answers. We use this example, drawn from a reform-based, epistemologically-aware classroom, to argue for the existence of a disconnect in ADK between physicists and physics students, even in ``best case'' scenarios like Stu's class.  We frame this paper as an existence proof of what can be learned about students' ADK from their problem solving, not as an exhaustive study into all the things students do with screencasts.

\section{Theoretical Framework}
We use ADK (outlined above) in concert with problem solving to study students' behaviors and values in physics classrooms. In the past there have been several discussions about the importance of teaching students the proper methods to solve problems. There are classes that are based on problem-based learning\cite{Etkina2010,Kuo2013Blending}, because studies have shown that solving problems develops students critical thinking skills, which in turn helps them develop a better understanding of the concepts behind the problems.

Cognitive theories describe the problem-solving process in four steps:\cite{Maloney1993a,Hsu2004} (1) understand the problem; (2) devise a plan; (3) carry out the plan; (4) verify or evaluate solution. Of course, these steps may be recursive for large or complex problems. Good solving-problem skills are an important part of ADK for both physicists and students. However, the kinds of problems that students solve are simpler, requiring few sub-problems or recursions. Physicists believe that the process of solving problems in physics classes is necessary and important practice for future professional situations. In contrast, most students seem to believe that the process of solving problems is a requirement for them to do well at school, just another step they need to get through. Students epistemological beliefs affect how and what they learn\cite{Hammer2003a}, as well as their future professional aspirations\cite{Packard2003}. In this paper we will discuss how discrepancies in students' and teachers' ADK may also have an effect on the student learning process.

\section{Context}
 This research is part of a larger project that was conducted in five high school Physics classes (four regular classes and one advanced class) in an urban area of a U.S. Western State.\cite{VanDusen2014FromFear,VanDusen2014dissertation,Nicholson-Dykstra2013,VanDusen2012Influencing}  The school was in a lower socio-economic status community which was primarily composed of students from non-dominant backgrounds. Table \ref{tab:schooldemo} shows the school demographics.  Minority groups composed the majority of the population in the school. Average student enrollment in physics was around 140 students, in classes with an average enrollment of 35. Most physics students were juniors. The remainder were sophomores. This study focuses on a classroom artifact from a single AP physics class that was composed of 32 juniors and seniors. 
 
 \begin{table}[hbp]
\caption{School Demographics\label{tab:schooldemo}}
%\vspace{0.15 in}
\begin{tabular}{|c|c|c|c|}
\hline
\multicolumn{4}{|c|}{Ethnicity} \\
Hispanic & White & Asian & Afr. Am.\\
56\%&	32\%&	8\%&	3\% \\
\hline\hline
\multicolumn{2}{|c|}{ Free or reduced lunch} & ESL/FEP & IEP\\

\multicolumn{2}{|c|}{41\%}&	49\%&	11\% \\
\hline
\end{tabular}
\end{table}
 
 The class, like all physics classes in this school, was taught by a teacher with a background in biology, including a Ph.D. in biochemistry. Like the majority of US high school teachers of physics\cite{Hodapp2009}, the teacher did not have a degree in physics. The teacher was recruited for the research project through her role as a teacher leader in a teacher-driven professional development community called Streamline to Mastery.\cite{VanDusen2012Changing,Ross2011Teacher} As a member of the program, the teacher aided in researching her own classroom practices in addition to performing her own action research projects.\cite{Nicholson-Dykstra2013} Because of the class' participation in this program, the classroom was outfitted with enough iPads to have one for every student, a technological affordance which dramatically changed classroom practice\cite{VanDusen2014}.

%Add a few lines here about the other artifacts (notebook) students produced. How were these different or alike to the screencasts

%A variety of data streams were collected during the school year, including field notes, video recordings/screencast and students notebook problems. As part of their grade students had to solve a set of problems weekly in a notebook that was handed in at the end of each week to be graded. Of the set of problems to solve in the notebook one had to be recorded as a screencast to be shown to the rest of the class. In this paper, present a case of a single 

%%%%%%%%%%%%%%%%%%%%%%%%%%%%%
\subsection{Screencasts}
As part of the class requirements, students had a notebook where they solve problems selected by the teacher. In addition to solving traditional end-of-chapter problems in their notebooks and worksheets, students selected one of their assigned problems for screencasting and sharing with classmates. Screencasting is a technology that captures a video of the iPad's screen while using the microphone to capture and merge audio to that file. This technology allows students to create various types of dynamic presentations. Within the physics class, the students' assignment prompts were to use screencasts to create a tutorial teaching how to solve a problem of their choice from a specific worksheet or chapter on their book. How these tutorials were to be created and what they should look like was not specified by the teacher and was left for the students to determine.  The students posted their screencasts on the class' Edmodo\texttrademark site (a Facebook-like social media site created specifically for education). After students posted their screencasts online, the teacher prompted students to view several peers' work and to provide their peers with feedback on the quality and usefulness of their screencasts.  Through viewing their peers' work and commenting on it -- both online and in class -- the students developed norms for what constituted ``good'' screencasts.\cite{VanDusen2012Influencing} 

Typical screencasts consisted of students talking through their thought process for solving a problem while writing out the solutions in real-time or while showing specific parts of a solution they had written down prior to recording the screencast. Because the screencasts record the display of the iPad, they do not include the students' hands or gestures; to draw their classmates' attention to specific parts of the screen, students often circle or underline with emphasis. 

 Each screencast is a ``presentation" that the students create for their peers.  Students sometimes use funny voices or altered inflection to indicate that they are reading the problem statement. In this analysis, we are not concerned with the production value of the screencast, but with what the student is trying to communicate. We draw on the screencast as a source of evidence about what the student values and perceives as ``doing well'' when solving a physics problem. 

Even though the practice of making and sharing screencasts is novel, we contend that the practice of solving this kind of problem in this kind of method is deeply typical of US physics education: it's very well-aligned with how the AP physics test assigns credit (and subsequently the teacher gives credit).  It's also well-aligned with prescriptive models of problem solving (e.g. \cite{Heller1992a}). The screencast affords us a glimpse of the ADK evinced in problem solving. Because the screencasts were created by the students in their own time, there is no record of students initial reactions to the problem and how they tackled it.  This performative, reflective aspect of their practice amplifies students' behaviors around ``good'' problem solving, because they are highlighting for their peers what they believe it is important in the process.

\section{Research Methods}
For the purpose of this study we chose to use case-based research method to study students' ADK through screencast. The decision of choosing case-based research was due to the fact that we are producing an in-depth descriptions and interpretations of a student's screencast solution of a physics problem to establish possible causal links between students' belief of what counts to do well in physics and their performance while explaining his peers how to solve a physics problem. Whilst it will not answer a question completely, it will give some indications and allow for further elaboration on the topic of ADK in classroom behaviors and the goals of problem-solving skills.

To evaluate students beliefs of doing well; i.e. what student perceived as ADK we coded the screencasts students created for the class to classify students behaviors. The type of behaviors we were interested in were confidence level, signs of reasoning and understanding, such as language used to explain the solution of the problem, the degree of detail on the steps to solve the problem and the reasoning and sense making while explaining the solutions. 

Because we are interested in how students' ADK may be visible in their classroom behaviors, for the purpose of this paper we selected a screencast where there were interesting disconnects between Stu's professed values in problem solving (e.g. getting the right answer, using the right method) and his performance of the problem in the screencast.  In resolving these disconnects, Stu makes parts of his ADK apparent. Even though we have chosen this particular screencast for this paper, other screencast that were analyzed presented similar behaviors.
%%%%%%%%%%%%%%%%%%%%%%%%%%%%%

The steps taken to do the analysis began with several group viewings of the screencast. In these viewings we discussed our interpretation of the screencast, differences of opinions regarding tone of voice or student's attitude. For example, there was a debate on whether the student sounded surprise or not when his answer did not match the book's answer. After reviewing and discussing it several times, we reach an agreement. Once all differences were ironed out, the screencast was transcribed and each researcher conducted individual viewings of the screencast. Through the use of micro-genetic analysis\cite{Parnafes2007}, each researcher individually examined Stu's problem solving discourse\cite{Gee2000} as he solved a specific physics problem about air pressure. Stu's expectations for ADK were investigated through examination of his language use, prosody, and tone. Because of the nature of screencasts, we couldn't follow his body language; however, we also examined his ``screen language'': the ways he connected his argument visually and drew emphasis to particular parts of the screen. The screencast was viewed many times by all researchers. 

The individual analysis was followed by group consensus building, which led to the creation of a generative coding scheme. Generative coding is the process of creating codes according to salient features of the data (as opposed to \textit{a priori} codes). The application of these codes to the screencast created the empirical foundation for our interpretation and claims about the student's ADK. Through multiple viewings, assisted by the transcript, we developed emergent codes and claims about the events in the screencast.\cite{Jordan1995} Through discussion, we reached consensus about our interpretation, claims, and evidence.

\section{Analysis}
Our claim is that an element of students' ADK in physics class is to follow a defined procedure, which (if applied correctly) will certainly result in a correct answer. The answer can be verified in the back of the textbook. Following procedure is more important than conceptual understanding, critical thinking, or real-world reasonability of results, and the student is so focused on the procedure in his performance that he neglects other important details. This ADK is in stark contrast to professional physicists' ADK about problem solving and also to the teachers' intended ADK for the students. We support our claim with observations from the screencast. 

Stu selects a problem about air pressure for his presentation. In the problem, the students are to calculate the force on a sheet of paper measuring $20cm$ by $30cm$ due to air pressure, given an air pressure of $1.013\times10^5Pa$.  Unknown to Stu, the published answer on the back of the book is wrong.  It reports $6.8\times10^3N$, where instead it should report $6.08\times10^3N$.  

As it was mentioned above Stu had previously solved the problem and is creating the screencast with the purpose of ``teaching" his classmates how to solve that problem.

\begin{figure}[htp] 
\begin{center} 
\includegraphics[width=\figwide]{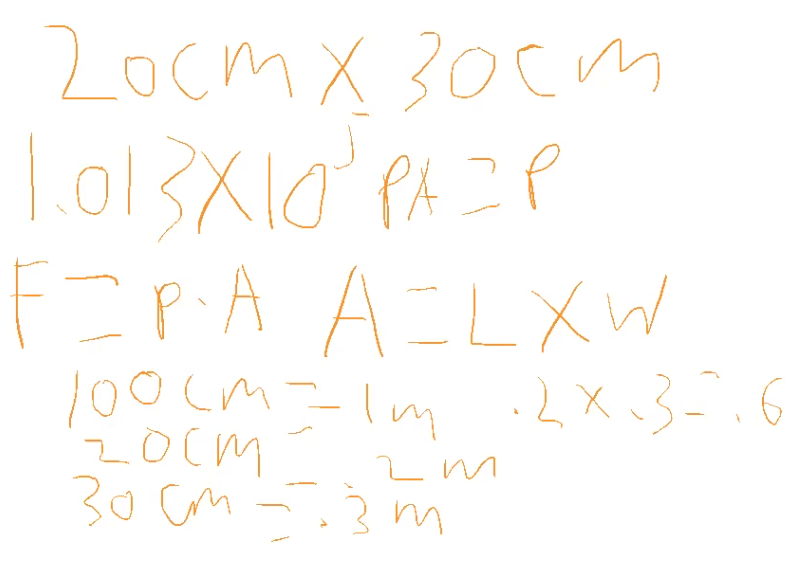} 
\caption{(Color) The first few steps in Stu's problem solving: stating what is known and setting up equations. The first two lines state the givens, the third line states general case equations, and the remaining lines are his calculation for the area of the sheet of paper. \label{fig:step1} } 
\end{center} 
\end{figure}

\begin{figure}[htp]
\begin{center} 
\includegraphics[width=\figwide]{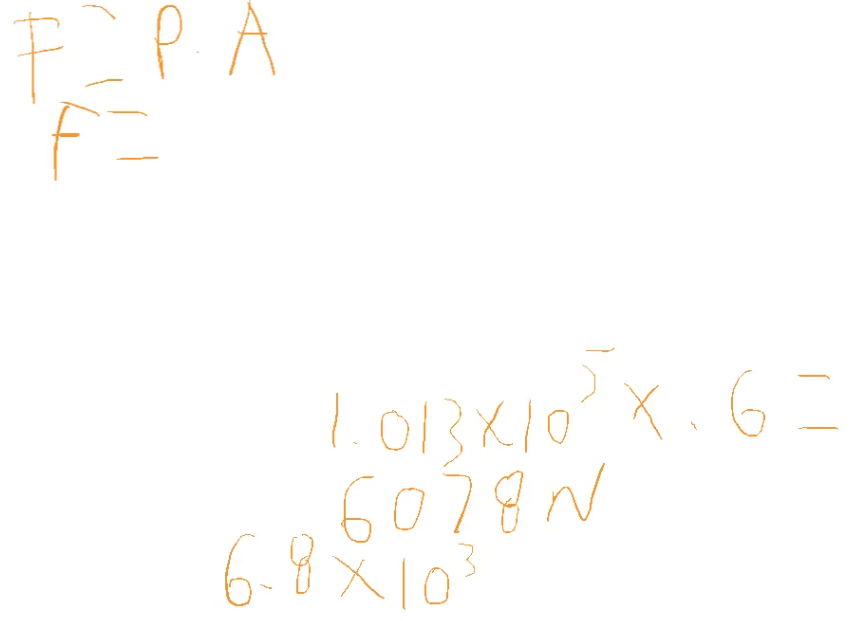} 
\caption{(Color) Stu's calculation of the force.  Stu lingers on this screen for a while, narrating how this answer does not match his expectations. \label{fig:step2} } 
\end{center} 
\end{figure}

\subsection{Overview of student's solution}
As he presents his solution, he follows the classical steps of problem solving, writing first the screen in Figure \ref{fig:step1} and then the one in Figure \ref{fig:step2}: 
\begin{description}
\item[Stating what is known] He writes on the iPad the variables the problem provides
\item[Define the unknowns] 
 In this case it is explicitly given on the problem, because it asks for the force. Although once he starts performing calculations, he identifies another unknown, which is the area he will need to calculate, in this case the force.
 
\item[Setting up equations]
Stu writes $F = p \times A$ No explanation is given about why he decided to use that particular equation or what the equation and its parts mean, other than the fact that is necessary. They were given an area and a pressure and learned that $F= p \times A$. Once he writes the equation for force he points out the variable pressure that was given and he moves on to say that they need the area and writes the equation for area as width x length. Again, no explanation as to why that quantifies the area.

\item[Performing calculations]
In the calculation part he is careful to write on the screencast the operations. He points out that before calculating the area he needs to transform from cm to m the variables of width and length. He does not comment on why he needs to change from cm to m, but he goes on detail about the operational way of doing it, as with the other calculations.
\end{description}

From what we observed, it was more important to him to teach and communicate to his peers the operational way of doing the steps to solve the problem, as opposed to the reasons why those steps were taken. That behavior agrees with our claims about the student's ADK. Even more evidence of it comes when he reaches the answer and there is a discrepancy between his answer and the book's answer. 

 Throughout the problem, he emphasizes that the steps he is taking to solve the problem are the correct procedure, but does not explain the conceptual reasoning behind his choices, such as why it's necessary to convert to SI units. The number that he produces at the end (Figure \ref{fig:step2}) is not the same as the one reported in the back of the textbook. One difference is that the book's answer is in scientific notation and has fewer significant figures, but we do not know if this is one of the facts that he sees as a problem or that the numbers even in the correct notation do not match. 
 
 \subsection{The student evaluates his work}
 At the end of his screencast, Stu says:

\begin{description}
\item[Stu] I checked with (mumbled) to see what answer we were supposed to get and it's $6.8\times10^3$, so I do not know about that. You just saw what I did here, so I guess it is possible I just mess something up\dots But all I know is that\dots Yeah, all I know is that this are the numbers I got. I followed the equations, and this is the answer I came up with here 6078 N. But according to this is supposed to be $6.8\times10^3$N.
\end{description}

He states that he cannot understand the discrepancies between his answer and the book's answer. He emphasizes that he followed the procedure, therefore his answer must be correct. He never mentions the difference in notations or shows a reasoning process for making sense between his answer and the book's answer. He does not show any evidence of valuing real-world reasoning or critical thinking; however, he does reassure his audience (his peers) that if they follow the procedure correctly, they will get the correct answer. From this we infer that following procedures correctly to obtain correct answers (which can be found in a trusted authority source) is an important part of ADK, but real-world reasoning or sense-making are not.

A reasonable counterargument might be that the student was solving a problem for the first time, in real time, for the screencast. If he didn't get a chance to prepare his answer, his solution might be choppy. However, we notice that the student's performance in the screencast could not have been his first time solving the problem. In presenting the problem, he mistakenly reports an area as $0.6cm^2$ instead of $0.06cm^2$. However, in following steps, his force calcuations use $0.06cm^2$. Also, we noticed that he does not do a self-check, not even to verify his procedure, which he seems so confident of. This could be because he had already solved the problem and tried to revise it before doing the screencast and could not find an error.

Additionally, his voice does not sound surprised or altered when reporting the book's answer to be different than his. This also indicates that he is not reasoning in the moment, but just doing mechanical steps because he cannot reconcile the idea of his answer being incorrect even though he followed the appropriate procedures.

The student values the correct procedure as the way to get the correct answer, and he advises his peers to use that procedure, even preferring it over his own answer. Again it is clear that what he thinks is important to teach to his peers is the steps needed to take in order to get the answer, not the reason for taking those steps.

\begin{description}
\item[Stu]
So, if you see a mistake in this process here, go on and change it to try to get this answer here\dots I guess\dots I guess I don't see why this is the answer, but\dots Try to get $6.8\times10^3$ using this method, correcting wherever I made my mistake. And if my answer is the right answer that is how you get it, but it's not though. Aim for this, but hopefully this process will help everyone.
\end{description}

It seems like the student has a very clear idea of the procedure: he seems confident on what he is doing and the steps he needs to take to solve the problem. This is what counts for him as doing well; this is his ADK for a physics class. 

Furthermore, as was mentioned earlier, as part of the class students had to complete sets of problems in their notebooks and one of the problems as a screencast. In both assignments, the teacher never formally graded the solutions. In a previous study screencasts and notebooks were scored for correctness and completeness\cite{VanDusen2014dissertation}. In this paper results show that when solving problems in screencasts students solutions were 21\% more complete then when solving problems in their notebooks. At the same time the students answers in the screencasts were 33\% more likely to be correct than when solving the notebook problems. This corroborates our claim, since it shows that for students it was more important to detail steps and procedures when teaching their peers than when just solving a problem for the teacher to evaluate.

\section{Discussion}

In this paper, we present an existence proof that students' ADK about problem solving is both visible in their problem solving discourse and substantially different from that of physicists.  On the one hand, this is not a revolutionary result: we already ``know'' that solving end-of-chapter style exercises\cite{Heller1992a} is a different kind of thing than conducting physics research.  However, through this work we bring forth the idea of ADK as a useful theory for analyzing what counts as doing well as a function of what kind of physics (e.g. high school classrooms) we study.

To make a more solid conclusions about problem-based learning and changes in ADK we would need to test different types of problems and control for variables, such as students' reactions when they first read the problem and think about the solution, because these screencasts are a finalized version of the solution that the student has already edited. This will provide with more information about the thinking process of the students before embarking in the problem-solving procedure and a less mechanical response. 

Future work could include involve interviews with students to establish what students are thinking and how they approach the problems initially.  However, the performative aspects of interviews are substantially different\cite{Russ2012} than for these ecologically valid screencasts, and we anticipate that aspects of students' discourse will change in the new setting.

As a technology for education, screencasts (and iPads more broadly) are important new research tools because they can give insight into how students understand physics and communicate it to their peers.  In traditional pen-and-paper problem solving, in which researchers have access only to students submitted inert solutions, screencasts present a new lens into the processes of students' problem-solving and communicating.   In this sense, they are the polar opposites of online homework systems.

Pedagogically, it is an open question as to whether screencasts improve student learning, or whether this kind of problem solving is one we'd like to promote.  In related work, we show that screencasts increase students' sense of ownership\cite{VanDusen2012Influencing,Nicholson-Dykstra2013} by acting as boundary objects, which is a valuable learning goal in and of itself. In a related data set from upper-division physics, students who complete more homework problems in pencasts (a close technological relative to screencasts) are more successful in the remainder of their work for the class\cite{WeliweriyaPencasts}.

\section{Acknowledgments}
The authors are indebted to insightful colleagues who offered criticism on this work (in alphabetical order), Paul Irving and Valerie Otero.  We are also grateful to the instructor who shared her class with us.  

This work was partially supported by the KSU Department of Physics and the National Science Foundation (DUE-934921).  Any opinions, findings, and conclusions or recommendations expressed in this material are those of the authors and do not necessarily reflect the views of the National Science Foundation or other funders.

%\bibliography{/Users/le/Dropbox/Research/Bibfiles/library}

\end{document}